\documentclass[12pt]{iopart}
\def\0{{\sst{(0)}}}
\def\1{{\sst{(1)}}}
\def\2{{\sst{(2)}}}
\def\3{{\sst{(3)}}}
\def\4{{\sst{(4)}}}
\def\5{{\sst{(5)}}}
\def\6{{\sst{(6)}}}
\def\7{{\sst{(7)}}}
\def\8{{\sst{(8)}}}

\newcommand{\be}{\begin{equation}} 
\newcommand{\ee}{\end{equation}}
\def\ba{\begin{array}}
\def\ea{\end{array}}

\def\sst#1{{\scriptscriptstyle #1}}

\newcommand{\bea}{\begin{eqnarray}} 
\newcommand{\eea}{\end{eqnarray}}

\begin{document}

\title{The scientific work of Sonia Stanciu}

\author{Constantin Bachas
}

\address{ Laboratoire de Physique Th{\'e}orique de l' Ecole
Normale Sup{\'e}rieure \\
24 rue Lhomond, 75231 Paris cedex, France}

\begin{abstract}
The Leuven workshop on the `Quantum Structure of Space-time and the 
Geometrical  Nature of the Fundamental Interactions'
had a special session
dedicated to the memory of Sonia Stanciu. This is the  summary of
a talk delivered by the author on this  occasion.
\end{abstract}

%Uncomment for PACS numbers title message
%\pacs{00.00, 20.00, 42.10}

% Uncomment for Submitted to journal title message
%\submitto{\JPA}

% Comment out if separate title page not required
%\maketitle

%\section{Introduction}

  I was  asked by the organizers  to say a 
few words about the scientific life  of Sonia Stanciu, a young string 
theorist and  active member of this European Network, who passed away a 
few months ago after a short but devastating illness.

    Sonia started her scientific  career  in Bonn, where she arrived from her
native Romania  to work on a doctoral degree under the supervision of
Vladimir  Rittenberg. Her thesis \cite{thesis} was on  
 hierarchies of integrable equations,     
in  relation with what are  nowdays refered to as  `old 
matrix models'.  These describe certain types of matter fields
living  on random triangulated surfaces, {\it i.e.}
interacting with regularized  2d quantum gravity. 
The models simplify considerably
in the  continuum  or `double scaling'   limit \cite{mm}, in which
their  partition and correlation functions 
can be obtained  as solutions of differential equations. 
Matrix models thus raised the  hope
  of a non-perturbative approach to string theory, but this  
was quickly  frustrated by  two important and 
still open  problems: the  approach  is  
restricted to  bosonic
strings, and to  `spacetimes' with at most two dimensions.

Sonia studied supersymmetric extensions of the KdV hierarchy
as a possible road towards lifting the above restrictions. Her thesis 
begins with Scott-Russel's remarkable description  
of his first  encounter with  a solitary wave in
 the Union canal \cite{jsr}  and
continues with  a nice  introduction to 
the history of classical integrability. She  then goes on to describe her
own research,  initiated  in  two  articles  with Jose
Figueroa-O'Farrill \cite{skdv}. Jose was in fact the real scientific mentor
of Sonia's thesis. He would also  become her life-long collaborator,
and  her husband. 
 
  Sonia spent  the last  year of her graduate studies visiting
  Queen Mary College in London, and  then moved on to a
 postdoctoral position  
at  the ICTP in Trieste. During this time she continued to collaborate
 with Jose,
first on possible obstructions to  gaugings  of Wess-Zumino-Witten
 models \cite{wzw},
and then on $\sigma$-models based on non-semi-simple Lie algebras \cite{nappi}
which   generalized   the Nappi-Witten plane-wave background \cite{Nap}. 
As with  most of their other joint work, these articles were characterized by
mathematical precision  and by a special affinity for algebraic
structures.
They  prepared the ground for the next class of problems  that
Sonia attacked, now as a mature  physicist: the embedding of
D-branes in exact curved-spacetime backgrounds. This is the subject on
which she had the highest impact, and the one on which she
worked until the end.

    String theory predicts two kinds  of quantum corrections to 
classical geometry:
string-loop corrections and $\alpha^\prime$ corrections. Though dualities may
in some cases  exchange the  two, the $\alpha^\prime$ effects are  more
accessible thanks to  the powerful algebraic techniques of conformal
field theory. This is illustrated  by the success of 
boundary CFT in describing the D-branes of certain exact (super)string
backgrounds. In the semiclassical regime, D-branes are submanifolds
which may  carry a non-trivial gauge connection, and are solutions of the 
Dirac-Born-Infeld equations \cite{Dai}. 
These equations, together with an
infinite series of $\alpha^\prime$ corrections are, on the other hand, 
equivalent to  the condition of
conformal invariance on the  boundary of a string worldsheet, 
\be
T_{zz} = T_{\bar z \bar z} \ \ \ {\rm at}\ \ \ {\rm Re} z = 0 \ . 
\label{tt}
\ee     
Here $T_{ab}$ is the energy-momentum tensor, and (\ref{tt})
says that the net flow of  energy to the boundary of the half-plane
must be zero. 
The  point  is that in  many exact
CFTs,  such as  minimal models  or WZW models, 
one can bypass  the geometric intuition  
and  solve the above equation algebraically. 
 In this approach one must
supplement the  condition (\ref{tt})  with certain consistency or
 `sewing' constraints \cite{Car,Lew,Aug}  
that translate  the locality of the  
underlying field theory.

  Sonia was among the first to realize that this was a very fertile
and worth-pursuing  direction.  I remember listening to  her
talk  at a  meeting of Dutch particle theorists, at NIKHEF
in the Fall of 1999.  She  was moving from Imperial College to 
the Spinoza Institute in Utrecht at that time, and was invited to present her 
results. I came out  thinking that these were interesting, but that
I would not want  to work on the  subject myself. Surely  enough, 
this is what I was doing   for the next  couple of years. It  gave
me the opportunity to discuss with Sonia on many occasions, 
at conferences, via email,  and during  the Fall 2000  semester 
at the Institut Henri Poincar{\'e}.

  I will not attempt  to summarize  the applications of BCFT,  and
the interplay of geometric and 
 algebraic techniques,  in  the study of curved-spacetime
D-branes in string theory. There  exist already   many 
complimentary  reviews on this  subject [12--16]. 
Here I will limit myself    to some
comments on Sonia's contributions in the field. 

The simplest but
rich example of quantum geometry, which  has served as  prototype for
many studies,  is the WZW model based on the 
manifold of SU(2). The algebraic construction of  D-branes in this model
has been
known essentially since the pioneering work of Cardy \cite{Car}, while their
geometric interpretation was  only 
clarified in recent years \cite{Ale,Juerg,Bac}. The algebraic construction
proceeds by first imposing the  symmetry-preserving
boundary conditions on the affine currents~:
\be 
 J^a_z = - \Omega^a_{\ b} J_{\bar z} \ \ \ {\rm at}\ \ \ {\rm Re} z = 0 \ , 
\label{jj}
\ee
where $\Omega$ is an automorphism of the  SU(2) algebra. 
Because of  the Sugawara form of the energy-momentum tensor, 
eqs.~(\ref{jj})  imply   automatically eq.~(\ref{tt}).  
The sewing constraints
can  then   be solved by  Cardy's prescription, valid 
for   any rational CFT. In the semiclassical 
regime these Cardy states correspond to D-branes which 
wrap spherical  conjugacy classes,  and 
are stabilized against shrinking by a non-trivial magnetic flux. 

    In a series   of articles
Sonia tried  to extend these considerations in several 
directions~:   to general compact semi-simple Lie groups \cite{s4}, 
to  N=2 coset models \cite{s1},   to the Nappi-Witten plane-wave
background \cite{s2},  and to the WZW model based on the
non-compact group SL(2,R) \cite{s3}. Each of them 
presents  technical and conceptual  challenges,  which are still 
the subject of active research today (see for instance the talks of 
Angelos Fotopoulos, Lennaert Huiszoon and 
Marika Taylor  in these proceedings). 
The SL(2,R) WZW model is especially
interesting because it is part of  the near-horizon geometry  of
the stringy NS5/F1 black hole \cite{Mas}. Sonia was the first to study
the supersymmetric D-branes of this background   from the 
CFT point of view. We  continued her  work with
Marios Petropoulos, adding
a new type of D-brane with anti-de-Sitter geometry to 
her list \cite{pet}. These new branes span  twisted conjugacy classes of
SL(2,R), and are obtained when $\Omega$ is a  non-trivial 
outer automorphism. The extension of the 
argument in \cite{Ale} to the case of an outer automorphism  was 
in fact also one 
of Sonia's nice  remarks \cite{s4,Juerg}. 
The dual holographic description of these  D-branes, and the
construction of the corresponding  boundary states,
are challenging  problems that are still only partially resolved 
(see for instance \cite{boer,pon} and references therein). 

One of the last problems to interest Sonia was the  
definition of  D-brane charge. In the  SU(2) WZW model
   this  takes values in the finite
group $Z_{k+2}$ ,  where $k$ is the level of the current algebra. 
Roughly speaking, the charge is the magnetic flux through the D-brane,  
 defined modulo   large gauge tranformations of the 
Neveu-Schwarz $B$-field (see \cite{phy,phy1} for  various related 
`physicist' arguments). 
A mathematically more precise formulation,  in terms of relative
cohomology,  was given  in the last paper of Sonia with Jose
\cite{s5}  (see also  \cite{Kli,Gaw,Fre} for earlier  work).
In this approach one considers topological 
obstructions to defining the  $\sigma$-model on open worldsheets. 
With hindsight,   the appropriate  mathematical framework in which to define
the invariant  D-brane charge 
is  most likely  some  twisted version of K-theory \cite{Kth}. 
It is,  however, not yet  clear how  this ties together  with the 
boundary-state CFT approach, even for the next example in the list,
{\it i.e.}
the WZW model on the group manifold of
SU(3) \cite{fs,phy1,s6}. Sonia might have talked
about this today,  had illness not cut her life so short.

\vskip 0.2cm 
  Most of us will remember Sonia as a quiet colleague, sharing a real
passion for the  field to which she has contributed in her own
discreet  and elegant way. She will be greatly missed in this and in all 
future network  meetings.

%\bea  \label{ppwvmet}
%ds^2 &=& 2 dx^{+} dx^{-} + \sum_{I=1}^{8} 
%( dx^{I} dx^{I} - \mu^2 (x^I)^2 (dx^{+})^2), \\ 
%F_{+1234} &=& F_{+5678} = 4 \mu. \nn
%\eea

\end{document}